\documentclass[12pt]{article}
\usepackage{graphicx}
\usepackage{epsfig}
\begin{document}
\baselineskip=5.5mm
\newcommand{\be} {\begin{equation}}
\newcommand{\ee} {\end{equation}}
\newcommand{\Be} {\begin{eqnarray}}
\newcommand{\Ee} {\end{eqnarray}}
\def\lg{\langle}
\def\rg{\rangle}
\def\a{\alpha}
\def\b{\beta}
\def\g{\gamma}
\def\G{\Gamma}
\def\d{\delta}
\def\D{\Delta}
\def\e{\epsilon}
\def\k{\kappa}
\def\l{\lambda}
\def\L{\Lambda}
\def\om{\omega}
\def\Om{\Omega}
\def\t{\tau}
\noindent
\noindent
\begin{center}
{\Large
{\bf
Dynamic Kerr effect responses in the Terahertz-range
}}\\
\vspace{0.8cm}
{\bf Uli H\"aberle and Gregor Diezemann} \\

Institut f\"ur Physikalische Chemie, Universit\"at Mainz,
Welderweg 11, 55099 Mainz, Germany
\\

\end{center}
\vspace{1cm}
\noindent
{\it
 Dynamic Kerr effect measurements provide a simple realization of a nonlinear experiment. We propose a field-off experiment where an electric field of one or several sinusoidal cycles with frequency $\Omega$ is applied to a sample in thermal equilibrium. Afterwards, the evolution of the polarizability is measured. If such an experiment is performed in the Terahertz-range it might provide valuable information about the low-frequency dynamics in disordered systems. We treat these dynamics in terms of a Brownian oscillator model and calculate the Kerr effect response. It is shown that frequency-selective behaviour can be expected. In the interesting case of underdamped vibrational motion we find that the frequency-dependence of the phonon-damping can be determined from the experiment. Also the behaviour of overdamped relaxational modes is discussed. For typical glassy materials we estimate the magnitude of all relevant quantities, which we believe to be helpful in experimental realizations.
}

\vspace{0.5cm}


\vspace{0.5cm}

\noindent

\vspace{1cm}
\noindent

\section*{I. Introduction}
Dynamic Kerr effect studies have been utilized to monitor the reorientational dynamics of supercooled liquids for a long time \cite{dejardinbuch}. In such experiments, the time-dependent polarizability is observed after an electric field has been switched on or off ('field on' and 'field off' version). In the context of supercooled liquids a comparison of reorientational correlation times extracted from Kerr effect data and dielectric relaxation showed that the correlation times associated with different Legendre polynomials ($l=2$ for the Kerr effect, $l=1$ for dielectric relaxation) are very similar \cite{williams}.

In a theoretical description of the dynamic Kerr effect, the coupling of the field to permanent dipole moments as well as to induced dipole moments has to be taken into account. Within a rotational diffusion model, a monoexponential decay ('field off') or a biexponential rise ('field on') of the polarizability has been found \cite{cole}.

An important property of the Kerr effect is its nonlinearity with respect to the applied field. Measurements of nonlinear effects generally provide information about physical properties that is not accessible via linear experiments. An example for this is the distinction between dynamic homogeneous and heterogeneous relaxation scenarios \cite{bamzeil}. A nonlinear experiment that was designed particularly in order to make that distinction is nonresonant holeburning (NHB)  \cite{roland}. Here, a large sinusoidal electric field is applied to a sample and the linear response of the nonlinearly perturbed system is measured afterwards. The idea is to specifically address those dynamical features or relaxational modes which absorb most of the energy supplied by the sinusoidal field. This of course depends on the frequency of the latter. Whether or not a frequency selective modification of the response is possible allows to distinguish between heterogeneous and homogeneous dynamics if the definition of reference \cite{bamzeil} is used for these terms.
 
Realizations of NHB on different systems in most cases lead to the conclusion that the dynamics are of a dynamic heterogeneous nature \cite{hetreviews}, but also dynamic homogeneous behaviour has been reported, e.g. in an ion-conducting glass \cite{sonesubstanz}. All NHB experiments performed so far utilized fields with frequencies in the kHz to Hz range, which is ideally suited to study slow relaxation phenomena in complex systems like supercooled liquids or spin glasses.

In a recent publication \cite{unsers}, we presented a model to describe an NHB-experiment on a much faster time scale, in the range of inverse Terahertz (THz). These short timescales are experimentally accessible since recent advances in the preparation of ultrashort laserpulses should allow to provide one or several coherent cycles of a sufficiently strong electric field in that frequency range \cite{keithnelson}. We found that in a Brownian oscillator (BO) model frequency-selective behaviour can also be expected for vibrational (fast) dynamics. 

Dynamic processes in the THz-range are mainly of a vibrational nature. In glasses, excess states compared to the Debye behaviour are found in the vibrational density of states (DOS), forming the so-called boson peak. Although there is consensus about the vibrational character of these dynamics, the lifetime of these states and the damping mechanism involved is still a point of interest \cite{handrich}.

Due to the fact that the Kerr effect signal is of a nonlinear nature, we expect Kerr effect measurements (in the THz-range) to be simpler to realize than NHB. Therefore, we study the Kerr effect response in this paper, particularly with regard to frequency-selectivity and its consequences. In contrast to the purely dissipative diffusion models which describe the dynamics on long timescales, we include inertial effects, implement a BO-model \cite{mukki} for the collective THz-dynamics and calculate the Kerr effect response function. The form of the signals we find allows to extract the frequency-dependence of the damping, and therefore provides important information about the vibrational dynamics.

The remainder of this paper is organized as follows. In the following section II we present the theory. We introduce our model in the first part, followed by a discussion about vanishing and nonvanishing contributions to the response in a second part. The latter point is deepened in a third part, where we give numerical estimates of the involved quantities. Section III contains the results and a discussion. We summarize and conclude in part IV.

\section*{II. Theory}

\subsection*{1. Nonlinear response functions in the BO-model}
We describe the dynamics of the system in the THz-range by a set of normal modes $\{ q\}=\{q_1, q_2,\dots, q_n\}$. In a molecular system the $q_i$ stand for internal degrees of freedom (intramolecular modes), but our discussion is focussed on macroscopic systems, especially glasses and disordered solids. In this case, the $q_i$ are normal coordinates for the collective vibrations around the boson peak (intermolecular and intramolecular modes). The classical Langevin equation for each mode in the BO-model reads as \cite{mukki}
\be \label{eqmotion}
m \ddot{q}_i+m \gamma_i \dot{q}_i+\frac{dV(\{q\})}{d q_i}=\G_i(t)-{\partial {\mathcal H}_{ext} \over \partial q_i}
\ee
Here, $\g_i$ is an Ohmic friction constant associated with the mode $i$, $m$ is a (mode-independent) mass, $\G_i(t)$ a Gaussian stochastic force with zero mean and correlation $\lg \G_i(t)\G_k(t')\rg =2\d_{i k}\g_i \b^{-1} m\d(t-t')$, $\b^{-1}=k_BT$, and $V(\{q\})$ the potential energy depending on all coordinates. For $V$ we choose the uncoupled anharmonic oscillator potential
\be \label{potential}
V(\{q\})=\sum_i V_i(q_i) = \sum_i m\om_i^2\left(\frac{1}{2}q_i^2+\frac{1}{3}\Theta_3 q_i^3+\frac{1}{4}\Theta_4 q_i^4+\dots \right)
\ee
where we scale the anharmonicity constants $\Theta_k$ by the harmonic eigenfrequencies. This means that the relative strength of anharmonicities with respect to the square of the corresponding harmonic eigenfrequencies is constant for all modes. In principle, we could include couplings between the different modes in the potential function, but we neglect these effects for a qualitative study and consider the system as a set of independent oscillators.

External electric fields couple to the system via a permanent dipole interaction and a polarizability interaction. Due to the fact that the permanent dipole moment is a first rank tensor and the polarizability is a second rank tensor, the expressions for the corresponding energy contributions are proportional to the first and second Legendre polynomials $P_l(\cos \vartheta )$, $l=1,2$. Here, $\vartheta$ is the angle between the direction of the applied field and the internal axis of the dipole moments. The energy contributions have the form $E^{perm}=\mu E P_1(\cos \vartheta)$ and $E^{ind}=\alpha E^2 P_2(\cos \vartheta)$ with the electric field $E$. The corresponding external Hamiltonian is thus given by
\be\label{exthamilton}
{\cal H}_{ext}=-\int\! d{\bf r}\left[ \mu({\bf r},\{q\})E({\bf r},t)\ P_1(\cos \vartheta) + \alpha({\bf r},\{q\})E^2({\bf r},t)\ P_2(\cos \vartheta)\right]
\ee
Strictly speaking, $\mu$ and $\alpha$ in this expression are dipole- and polarizability densities. In the following calculations, the applied electric field is chosen as $E({\bf r},\tau)=E*e^{i{\bf kr}}\sin(\Omega \tau)$ up to $t_p=2\pi N / \Omega$, and $0$ afterwards. Here, $N$ denotes the number of cycles applied, $E$ is the time-independent field amplitude, and ${\bf k}$ the corresponding wavevector. The field sequence is shown in figure $1$.  We define the time variable $t$ starting with $0$ at $t_p$ and we calculate the time-dependent evolution of the polarizability after the system has been driven out of equilibrium by the sinusoidal field, that is the expectation value
\be \label{definebs}
\lg\ \lg P_2(\cos \vartheta ) \int d{\bf r}\ \alpha({\bf r} ,\{q\}) \rg\ \rg_\vartheta
\ee
with the outer brackets $\lg\dots\rg_\vartheta$ denoting an orientational average.

We assume that external THz-fields do not affect the orientations $\vartheta$. This is because typically the reorientational dynamics in supercooled liquids occur on the timescale from microseconds to seconds in the temperature range we are mainly interested in. For completeness, however, we also present results for a rotational diffusion model in Appendix B.

The (spatially integrated) polarizability in eq.(\ref{definebs}) is calculated in perturbation theory with the external field as the perturbation. Its time-dependence is a consequence of the interaction with the time-dependent external Hamiltonian (\ref{exthamilton}), consisting of two terms describing permanent dipole moment and polarizability interactions. As a consequence of the fact that orientations are unaffected by the external field the linear response of order $E$ vanishes if isotropic systems are considered. This is because the only linear term in $E$ is the permanent dipole interaction term in (\ref{exthamilton}) which is proportional to the first Legendre polynomial. Together with the second Legendre polynomial in (\ref{definebs}) the contribution vanishes after performing the orientational average $\lg\dots\rg_\vartheta$ due to the Legendre polynomials' orthogonality.

The lowest nonvanishing order in $E$ is therefore ${\mathcal{O}}(E^2)$. In contrast to many other nonlinear experiments, e.g. NHB, where nonlinear contributions are observed as small variations on a usually large linear response background, all the obtained signal in Kerr effect measurements is of a nonlinear nature. In detail, the polarizability (\ref{definebs}) in ${\mathcal{O}}(E^2)$ has contributions $\propto \alpha \alpha$ from the induced dipole interaction term in (\ref{exthamilton}) treated in first order perturbation theory, and $\propto \alpha \mu \mu$ from the permanent dipole moment interaction treated in second order perturbation theory \cite{meinsenf}. The orientational dependencies of these terms are proportionalities to $P_2P_2$ and $P_2P_1P_1$, both nonvanishing in the orientational average.     

These terms are the contributions found in a rotational diffusion model \cite{cole}. In the BO-model defined by eq.(\ref{eqmotion}), the interaction with applied electric fields is given by ${\partial {\cal H}_{ext} \over \partial q_i}$. Therefore, derivatives of the dipole moments with respect to the $q_i$ replace the dipole moments. No external force is acting on the system if permanent and induced dipole moments do not depend on the coordinates $q_i$. This need of a coordinate dependency of the dipole moments for a coupling of external fields to the system yields a selection rule for infrared and Raman activity. For our calculations, we expand dipole moment and polarizability in a Taylor series in the $q_i$
\Be
\label{mutaylor}\mu({\bf r} ,\{q\})=\mu_0 ({\bf r})+\sum_i \mu'({\bf r})_iq_i +\sum_i \frac{1}{2}\mu''({\bf r})_i q_i^2 + \dots \\
\label{alphataylor}\alpha({\bf r} ,\{q\})=\alpha_0 ({\bf r})+\sum_i \alpha'({\bf r})_iq_i +\sum_i \frac{1}{2}\alpha''({\bf r})_i q_i^2 + \dots 
\Ee
where primes denote partial derivatives with respect to $q_i$, evaluated at $q_i=0$ for all $i=1, \dots, n$. For simplicity, we neglect crossterms ($\propto q_i q_j$) in both expansions. 

The Taylor-expansion of the polarizability (\ref{alphataylor}) allows to split the expectation value (\ref{definebs}) into a sum $B_0+B(t)$ with
\Be\label{bnull}
B_0 \!& = \! & \lg\ P_2(\cos \vartheta) \int \! d{\bf r}\ \alpha_0({\bf r})\ \rg_\vartheta \\ \label{bvont}
B(t)\! & = \! & \sum_i B_i(t)=\sum_i \lg\ P_2(\cos \vartheta ) \left(  \int \! d{\bf r}\ \alpha'_i({\bf r})\ \lg q_i(t) \rg+\frac{1}{2} \int \! d{\bf r}\ \alpha''_i({\bf r})\ \lg q_i^2(t) \rg +\dots \right)\ \rg_\vartheta \
\Ee
Apart from a prefactor, the quantity $B_0+B(t)$ is the so-called birefringence function. The prefactor is given by $2\pi /(Vn)$ with the sample volume $V$ and the mean refractive index $n$ if $\alpha$ is identified with the difference between the polarizabilities parallel and perpendicular to the molecule's symmetry axes \cite{deschardaeng}. 

$B_0$ in eq.(\ref{bnull}) does not depend on the normal modes of the system and is therefore time-independent in the BO-model. While $B_0$ is the relevant contribution if molecular reorientations are considered (as discussed in Appendix B), we have to deal with $B(t)$ in the BO-model. According to eq.(\ref{bvont}), the time-dependent expectation values $\langle q_i(t)\rangle$ and $\langle q_i^2(t) \rangle$ have to be be calculated, where the former leads to contributions proportional to $\alpha '_i$, and the latter to contributions proportional to $\alpha ''_i$. Therefore the Fokker-Planck equation corresponding to eq.(\ref{eqmotion}) has to be solved. 

The functional form of the expectation values basically is a consequence of the interactions with the external field. These interactions are treated in time-dependent perturbation theory. In the expression for the external Hamiltonian (\ref{exthamilton}) again the Taylor expansions of dipole moment (\ref{mutaylor}) and polarizability (\ref{alphataylor}) are used. From these expansions it is clear that different derivatives of the dipole moments enter the expressions for the expectation values $\langle q^n_i(t) \rangle,\ n=1,2,\dots$. As mentioned above, terms of order $E^2$ arise from the polarizability interaction treated in first order perturbation theory (eq.(\ref{qalpha}) below), and from permanent dipole moment interaction treated as a second order perturbation (eq.(\ref{qmumu})). The perturbation ansatz yields expectation values of the form
\begin{eqnarray} \label{qalpha}
\langle q^n_i(t)\rangle & = & E^2 P_2 \sum_{k=1,2,\dots} \int d^3{\bf r'} e^{i{\bf k r'}} \alpha^{(k )} ({\bf r'})_i \ f^{k}_i(n, \Xi_i,t,t_p)  \\ \label{qmumu}
\langle q^n_i(t)\rangle & = & E^2 P_1P_1 \sum_{k,l=1,2,\dots} \int d^3{\bf r'} \int d^3{\bf r''} e^{i{\bf k(r'+r'')}} \mu^{(k)} ({\bf r'})_i \mu^{(l)} ({\bf r''})_i \ f^{(k,l)}_i(n,\Xi_i,t,t_p)
\end{eqnarray}
Here, $n=1, 2,\dots$, $P_l=P_l(\cos \vartheta )$ and $(\alpha^{(k)}_i, \mu^{(k)}_i, \mu^{(l)}_i)$ denote the $k$-th or $l$-th order derivatives of $\alpha$ or $\mu$ as in the Taylor expansions (\ref{mutaylor},\ref{alphataylor}). Furthermore, $f^{k}_i(n,\Xi_i,t,t_p)$ and $f^{(k,l)}_i(n,\Xi_i,t,t_p)$ are some time-dependent functions. Here, $\Xi_i$ is an abbreviation for all the parameters that characterize the mode, namely $\omega_i, \gamma_i$ and the anharmonicities $\Theta_3, \Theta_4 ,\dots$. In calculating the functions $f^{k}_i$ and $f^{(k,l)}_i$, the difficulty arises that the solution of the Fokker-Planck-equation is only known for harmonic potential. Therefore, the anharmonicities $\Theta_j$ must also be treated in (time-independent) perturbation theory.

The appearance of the different derivatives and the anharmonicities makes the situation quite complex. In the following section, we discuss the relevance of these different contributions. That means we discuss which derivatives appear in the terms of order $\alpha \alpha$ and $\alpha \mu \mu$.

\subsection*{2. Contributions to the Kerr effect response}

In the BO-model, some of the lowest-order contributions vanish. Tables $1$ and $2$ list the lowest nonvanishing contributions for $\alpha\alpha$- and $\alpha\mu\mu$-terms. In this section, we skip the lower index $i$ of the derivatives.
\begin{table}[h]
\begin{center}
\begin{tabular}{|c|c|} 
  \hline
   contribution & order   \\
  \hline
   $\alpha' \alpha'$ & 0 \\
   $\alpha'' \alpha''$ & 2 \\
   $\Theta_3 \alpha' \alpha''$ & 2  \\
   $\Theta_3 \alpha'' \alpha''$ & 3\\
   $\Theta_4 \alpha' \alpha'$ & 2  \\
   $\Theta_4 \alpha' \alpha''$ & 3 \\
   $\Theta_4 \alpha'' \alpha''$ & 4 \\
 \hline 
\end{tabular}
\caption{Nonvanishing contributions to the response of order $\alpha \alpha$. A definition of the listed 'order' is given in the text.}
\end{center}
\end{table}
\begin{table}[h]
\begin{center}
\begin{tabular}{|c|c|} 
  \hline
   contribution & order  \\
  \hline
   $\alpha'' \mu' \mu'$ & 1 \\
   $\alpha' \mu' \mu''$ & 1 \\
   $\alpha'' \mu'' \mu''$ & 3 \\
   $\Theta_3 \alpha' \mu' \mu'$ & 1 \\
   $\Theta_3 \alpha'' \mu' \mu''$ & 3  \\
   $\Theta_3 \alpha' \mu'' \mu''$ & 3  \\
   $\Theta_3 \alpha'' \mu'' \mu''$ & 4 \\
   $\Theta_4 \alpha'' \mu' \mu'$ & 3 \\
   $\Theta_4 \alpha' \mu'' \mu'$ & 3  \\
   $\Theta_4 \alpha'' \mu'' \mu'$ & 4 \\
   $\Theta_4 \alpha' \mu'' \mu''$ & 4  \\
   $\Theta_4 \alpha'' \mu'' \mu''$ & 5  \\
 \hline 
\end{tabular}
\caption{As in Table 1 for the $\alpha \mu \mu$-contributions.}
\end{center}
\end{table}
The terms not listed $\propto \alpha ' \alpha ''$, $\Theta_3\alpha '\alpha '$, $\alpha ' \mu '\mu '$, $\alpha '\mu ''\mu ''$, $\alpha ''\mu '\mu ''$, $\Theta_3 \alpha ''\mu '\mu '$, $\Theta_3\alpha '\mu '\mu ''$ and $\Theta_4 \alpha' \mu ' \mu'$ all vanish identically in the BO-model, independent of the functional form of the applied field. This also holds if one or several of the $\mu^{(n)}$ are replaced by $\alpha^{(n)}$. The tables are valid for all Raman experiments or photon echoes, at least as long as crossterms in the potential and in expansions of the dipole moments are neglected.

We define a system to count the 'order' of the terms in the tables as follows. The leading terms in the forces (derivatives of ${\cal H}_{ext}$ with respect to the $q_i$) are $\mu '$ and $\alpha '$. These first derivatives are defined as order $0$, and each additional derivative increases the order by $1$. Finally, the order of all appearing derivatives is summed up. A similar system is used for the anharmonicities. Here, the parabolic potential describes the 'unperturbed' state while anharmonicities are considered as perturbations. If a term is proportional to the cubic or quartic anharmonicity $\Theta_3$ or $\Theta_4$, we therefore increase the order by $1$ or $2$, respectively. We argue that the lowest order terms in the tables are the most relevant contributions to the response, and thus assume that polarizability, dipole moment and potential have fast converging Taylor series. In order to justify this approach we have to compare $\alpha '$ to $\alpha ''$, $\mu '$ to $\mu ''$ and the cubic and quartic anharmonicities $\Theta_3$ and $\Theta_4$ (next section).   

 We further assume that terms where different order derivatives of the same quantity ($\alpha ,\mu$) appear can be neglected. This is because they appear in the form of spatial correlation functions as e.g. in equation (\ref{qmumu}). Note that this is a common ansatz in glassy systems, but the argument is not necessarily valid in molecular systems. Here the derivatives of dipole moments appear just as numbers, not in the form of correlation functions, and there is no general argument for neglecting a term $\propto \alpha '\mu '\mu ''$ compared to the term $\propto \alpha'' \mu' \mu'$, for example.

\subsection*{3. Estimate of the relevant quantities}

\subsubsection*{Polarizability and dipole moment}

The Taylor-expansion of the polarizability (\ref{alphataylor}) is often truncated after the linear term (Placzek-approximation) \cite{murrfourk}. The inclusion of the second derivative frequently appears in context with nonlinear $\chi_5$-spectroscopy. This is because there is neither a contribution $\propto \alpha ' \alpha ' \alpha '$ to the $\chi_5$ signal (see also \cite{okumurasehrcool}), nor a contribution $\propto \alpha ' \mu ' \mu '$ (see the table above). In an oscillator model, either anharmonicities or the second derivative of the polarizability are needed to find a nonvanishing contribution of second order in the permanent dipole moment.

In Kerr effect measurements on liquid water \cite{mukamel}, observed long lived exponential relaxations are inexplicable with a purely linear coupling of the field to the $q_i$. From comparisons of model calculations with the measured data, the ratio $\alpha '^2 : \alpha ''^2$ is found to be $50:1$.

A simulation of the dynamics in CS$_2$ has been analyzed in an instantaneous normal mode approach \cite{murryfourkas}. The tensorial nature of the polarizability is treated in more detail than in our scalar notation. Derivatives of the polarizability are calculated via finite differences. For the ratio $\alpha '^2 : \alpha ''^2$ (in our notation) values ranging from $(10:1)$ to $(100:1)$ have been found.
 
We therefore conclude that the second derivative of the polarizability can be neglected whenever it is directly compared to the first derivative. Nevertheless, $\alpha ''$ has to be taken into account for those terms where a replacement of $\alpha ''$ with $\alpha '$ yields zero due to the properties of the BO-model, see above.  

For the permanent dipole moment, typical values of the first derivative $\mu '$ are in the range of $1 Db/\AA$ \cite{harris}. An estimate of the second derivative $\mu ''$ is difficult. Here, we simply assume that the convergence behaviour of the permanent dipole moment's Taylor series is similar as for the polarizability $\alpha$.

\subsubsection*{Anharmonicities}

From a comparison of experimental values on a variety of glasses with the predictions of the soft potential model \cite{parshin} we find a mean value for the quartic anharmonicity of $\Theta_4 \approx 200 \AA^{-2}$. For an estimate of the cubic anharmonicity $\Theta_3$ we use thermal expansion, an effect untouched by $\Theta_4$ due to symmetry reasons. In a standard model \cite{kittel}, the average thermal displacement of an anharmonic oscillator described by a potential like in eq.(\ref{potential}) is $\lg x(T) \rg = k_B T \Theta_3/(m\omega_i^2)$. We estimate a temperature-dependent lattice constant (or atomic distance) $a(T)$ as $a(T) \approx a(T=0)+\lg x(T) \rg$. This leads to
\be \label{thermalexp}
\Delta a=a(T_2)-a(T_1)=\Theta_3 \frac{k_B \Delta T}{m\omega_i^2}
\ee
with $\Delta T=T_2-T_1$. From experimental data on solid argon \cite{kittel} (an fcc-crystal) we find an expansion $\Delta a\approx 0.1 \AA$ if the temperature is increased from $40 K$ to $80 K$ . The thermal expansion is linear in that temperature range. Divided by the average lattice constant, this results in a thermal expansion coefficient of $\approx 4.6*10^{-4} K^{-1}$. A reasonable value for $m\omega_i^2$ is $0.2 eV/\AA^2$, also based on \cite{parshin}. These values with the ansatz in eq.(\ref{thermalexp}) lead to $\Theta_3\approx 6\AA^{-1}$. Comparing now the corresponding terms in the equation of motion (\ref{eqmotion}), we find the ratio
\begin{displaymath} 
(m\omega_i^2\Theta_4 q^3) / (m\omega_i^2\Theta_3 q^2) =(\Theta_4 *q) / \Theta_3  \approx 30 * q/\AA. 
\end{displaymath}
Typical values for $q$ are $q\approx (0.01\dots 0.1)\AA$, being in the percent range of typical atomic distances in solids. Following these arguments, the quartic anharmonicity term is not necessarily smaller than the cubic anharmonicity term. Another estimate for $\Theta_3$ based on the soft potential model \cite{parshin} leads to the similar value of $\Theta_3\approx 15 \AA^{-1}$. This latter approach is doubtful because the two-level-systems should not be used for an estimate of a cubic potential anharmonicity, but all in all we conclude that the quartic anharmonicity is not necessarily small compared to the cubic anharmonicity. In our case, we would not neglect the term $\propto \Theta_4\alpha '\mu '\mu '$ compared to the term $\propto \Theta_3\alpha '\mu '\mu '$, but the latter term vanishes in the BO-model. The next nonvanishing order is $\propto \Theta_4\alpha ''\mu '\mu '$, which we neglect due to the relative smallness of $\alpha ''$ with respect to $\alpha '$. It should be mentioned that higher-order spectroscopy techniques are more sensitive to higher anharmonicities, as has been pointed out in ref.\cite{okumura7}.

\subsubsection*{Summary}

As a result of the proceeding estimates, we are left with the contribution $\propto \alpha '\alpha '$ and the terms $\propto \alpha ''\mu '\mu '$ and $\propto \Theta_3 \alpha '\mu '\mu '$. This finding is analogous to the conclusions from \cite{okumurasehrcool}, where the term $\alpha '\alpha '$ is found as the dominating part in the $\chi_3$ signal and corresponding terms $\propto \alpha ''\alpha '\alpha '$ and $\propto \Theta_3\alpha '\alpha '\alpha '$ are pointed out as the most relevant contributions to the $\chi_5$-signal. The former term $\propto \alpha '\alpha '$ is of minor interest to us, since it represents a quasi-linear response, apart from the fact that the field appears quadratically instead of linear. In our context we have a squared sine function in the time integral instead of the sine function we have in calculating the linear dielectric response ($\propto \mu '\mu '$). The analytical solution for the quasi linear term can be found in Appendix A. The other two contributions, denoted as 'harmonic' ($\propto \alpha ''\mu '\mu '$) and 'anharmonic' ($\propto \Theta_3 \alpha '\mu '\mu '$) terms in the following, are discussed in detail in the next section.

\section*{III. Results}

\subsection*{1. Solution for an individual mode}
 
Inserting the calculated (time-dependent) expectation values $\lg q_i \rg$ and $\lg q_i^2 \rg$ into (\ref{bvont}) we find harmonic and anharmonic contributions of the form
\Be \label{bharm}
B^{harm}_i(t)=E^2 \lg P_1P_1P_2 \rg_\vartheta \int d^3{\bf r} \alpha''({\bf r})_i \int d^3{\bf r'} \int d^3{\bf r''} \mu'({\bf r'})_i \mu'({\bf r''})_i e^{i{\bf k(r'+r'')}}\ f^{harm}_i(t,t_p) \\ \label{banharm}
B^{anharm}_i(t) =E^2 \Theta_3 \omega_i^2 \lg P_1P_1P_2 \rg_\vartheta \int d^3{\bf r} \alpha'({\bf r})_i \int d^3{\bf r'} \int d^3{\bf r''} \mu'({\bf r'})_i \mu'({\bf r''})_i e^{i{\bf k(r'+r'')}}\ f^{anharm}_i(t,t_p)
\Ee
We have set the mass in the Langevin equation (\ref{eqmotion}) to unity. The functions $f_i(t,t_p)$ are linear combinations of (complex) exponentials in time. The exact expressions are given in Appendix A. In order to see the general features of the response, it is instructive to consider the limit $t_p \to \infty$, when the system has reached a steady state before the field is switched off. For the harmonic term, we find
\be \label{fharmlimit}
f^{harm}_i(t,\infty)=\frac{1}{(\l_1-\l_2)^2}\frac{1}{2}\left(\frac{\Omega}{\l_1^2+\Omega^2}e^{-\l_1 t}-\frac{\Omega}{\l_2^2+\Omega^2}e^{-\l_2 t} \right)^2
\ee
where $\l_{1,2}=1/2(\gamma_i \pm \delta_i)$, $\delta_i=(\gamma_i^2-4\omega_i^2)^{1/2}$. Note that $\delta_i$ is complex for $\gamma_i < 2\omega_i$ (underdamped case, oscillatory motion), and real for $\gamma_i > 2\omega_i$ (overdamped case, relaxational motion). In the case of underdamped motion, the limit $t_p\to \infty$ is achieved if the number of applied cycles fullfills $N \gg \frac{\Omega}{\pi \gamma_i}$, while $N \gg \frac{\Omega}{2\pi} \frac{\gamma_i}{\omega_i^2}$ is required in the overdamped  case. By further assuming a strongly underdamped mode, terms of order $\gamma_i$ can be neglected in the denominators and the Fourier transform's imaginary part becomes
\be \label{approx}
f^{harm}_i(\omega,\infty)''\approx \frac{1}{2} \frac{1}{\omega_i^2} \frac{\Omega^2}{(\Omega^2-\omega_i^2)^2}\left[ \frac{1}{2}\frac{\omega}{\gamma_i^2+\omega^2} + \frac{\omega}{4\omega_i^2-\omega^2} \right]
\ee
We present the responses in the form of the imaginary part of the time signal's Fourier transforms since this representation is best suited to discuss the relevant features. This is due to the oscillating form of the responses in the underdamped case.

If linear response functions are calculated, a simple relation holds between the time-dependent response $R^{AC}(t)$ of the system to an oscillating field $E(t)=E \exp(-i\omega t)$ and the Fourier transformed response  $R^P(\omega)=\int_0^{\infty}dt\ e^{i\omega t}R^P(t)$ to a pulse field $E(t)=E\delta (t)$. This relation is $R^{AC}(t)=\exp(-i\omega t) R^P(\omega)$. It is valid in this form if the AC-field has been switched on at time $t=-\infty$. However, since we are dealing with nonlinear response functions here, no such relation exists \cite{unsers}\cite{hosokawa}. It is not possible to derive the Kerr effect response to the oscillating field from the pulse response. 

Expression (\ref{approx}) shows that the signal's amplitude becomes large if the pump-frequency $\Omega$ is close to the eigenfrequency $\omega_i$ of the (strongly underdamped) mode. This fact guarantees frequency-selective behaviour. In addition, signals in the Fourier transform's imaginary part ($\omega$-dependence) will appear at $\omega =2\omega_i$ and $\omega =\gamma_i$. The singularities in the function (\ref{approx}) stem from the fact that the damping has been neglected partially in deriving this expression. The expression for the anharmonic term is more complicated. It is not possible to give an expression as simple as eq.(\ref{approx}) for this term, not even in the limit $t_p \to \infty$. Similar as for the harmonic term, signals in the Fourier transforms' imaginary part appear at $\omega=2\omega_i$ and at $\omega =\gamma_i$ (due to the terms $\propto \exp(-(\l_i+\l_j)t)$ for $i=j$ resp. $i\ne j$ in the time signal, see Appendix A). Furthermore, we find a signal at $\omega=\omega_i$ from the terms $\propto \exp(-\lambda_i t)$. Figure $2$ shows examples of underdamped mode responses for both harmonic (upper panel) and anharmonic (lower panel) contributions. For the harmonic term, the discussed limit $t_p\to\infty$ is shown (dashed line) in addition to results for a single applied cycle (full line). For the anharmonic term, we show results of one cycle only for clarity. An increasing pump time $t_p$ in both cases mainly gives rise to a growth of amplitudes, apart from some variations in the relative intensities of the features at $\omega=\omega_i, 2\omega_i, \gamma_i$. We do not discuss in detail the differences between harmonic and anharmonic contributions here, although the different functional forms might allow to distinguish between the different origins of the nonlinear signal (see also \cite{okumurasehrcool}). We will focus on the common features in the signals. Whether anharmonic or harmonic contributions are of more importance will strongly depend on the system examined.

In addition to the underdamped limit, we consider the overdamped limit $\gamma_i \gg \omega_i$. This limit can be derived either from eqns.(\ref{bharm}) and (\ref{banharm}) by an expansion in powers of $\gamma_i^{-1}$ or by solving the equation of motion of the so-called Ornstein-Uhlenbeck process instead of the BO-model. Overdamped motion is relevant also in the Terahertz range, since comparisons of experimental data from Raman spectroscopy with theoretical considerations show that underdamped as well as overdamped response functions are needed to fit the data, like in liquid CHCl$_3$ and CS$_2$ \cite{okumurasehrcool}, or in CCl$_4$ \cite{tokmakoff}. We find
\begin{equation} \label{ouharm}
f^{harm, OU}_i(t,t_p)=\frac{1}{2}\frac{1}{\omega_i^4} \epsilon ''(\Omega)^2  (1-e^{-\Lambda_i t_p})^2 e^{-2 \Lambda_i t}
\end{equation}
and
\begin{equation} \label{ouanharm}
f^{anharm, OU}_i(t,t_p)=\frac{1}{\omega_i^6} \epsilon ''(\Omega) \left(\epsilon ''(\Omega) (1-e^{-\Lambda_i t_p})^2 e^{-2 \Lambda_i t} - 3 \epsilon ''(2\Omega) (1-e^{-\Lambda_i t_p}) e^{-\Lambda_i t} \right)
\end{equation}
where $\epsilon(\Omega)=\Lambda_i / (\Lambda_i- i \Omega)$ and $\Lambda_i=\omega_i^2/\gamma_i$. The appearance of the dielectric loss functions $\epsilon ''(\Omega)$ again guarantees frequency selective behaviour. Here, the maximum excitation is for pump frequency $\Omega =\Lambda_i$ and $\Omega =\Lambda_i /2$ (second term in $f^{anharm, OU}_i$), and signals appear at frequencies $\omega=\Lambda_i, 2\Lambda_i$. Overdamped mode responses are shown in figure 3 for harmonic (upper panel) and anharmonic (lower panel) terms. The curves are for one cycle of the pump field. In the overdamped case, the influence of additional cycles is very small, because the exponentials $e^{-\Lambda_i t_p}$ in (\ref{ouharm},\ref{ouanharm}) are very small after only one cycle if the pump frequency is in the range of $\Lambda_i$. The overdamped responses are very similar to results from a rotational diffusion model. The resulting expressions for this model are given in Appendix B for comparison.

\subsection*{2. Distribution of modes}

Having solved the problem for the individual modes, we now consider distributions of the parameters as a model for macroscopic systems. For the eigenfrequencies $\omega_i$ a Debye DOS $g(\omega_i)\propto \omega_i^m, m=2$ is a natural choice. We cut off the the distribution at the Debye-frequency $\omega_D$. Alternatively, a proportionality $g(\omega_i) \propto \omega_i^4$ ($m=4$) could be chosen as found in the soft-potential model, or a more realistic decay than the abrupt cutoff might be used. However, these details have little influence on our results, so we focus on the most simple-minded distribution.

In addition, the damping $\gamma_i$ in general is frequency-dependent. Among others, inelastic X-ray scattering techniques are capable to detect the dependence of sound attenuation on momentum transfer $Q$. Usually, a $Q^2$-law for the damping holds in the region of small $Q$ and breaks down at larger values of $Q$ \cite{sette}\cite{benassi}\cite{scopigno}. Together with a linear dispersion $\omega_i \propto Q$ in the region of small $Q$ we have a quadratic dependence of the damping on the eigenfrequencies, $\gamma_i(\omega_i)=b \omega_i^2$. Reasonable prefactors are in the range of $0.1$ if damping and frequency are in $meV$, as in glycerol $b\approx 0.12\ meV^{-1}$ \cite{sette}, in vitreous silica $b\approx 0.06\ meV^{-1}$ \cite{benassi} or in glassy selenium $b\approx 0.17\ meV^{-1}$ \cite{scopigno}. For glassy selenium, a validity of the relation $\gamma(Q)\propto \omega_i(Q)^2$ even beyond the linear dispersion region of $Q$ has been reported\cite{scopigno}.

Another important point is the $\omega_i$-dependence of the appearing dipole moment correlations in eqs. (\ref{bharm},\ref{banharm}). Here, we assume the same functional behaviour as found for the polarizability correlations, the so-called light-to-vibration-coupling $C(\omega_i)$. A detailed discussion of this ansatz is given in \cite{unsers}. As a consequence we approximately have 
\be\label{lightvib}
\int d^3{\bf r} \alpha''({\bf r})_i \int d^3{\bf r'} \int d^3{\bf r''} \mu'({\bf r'})_i \mu'({\bf r''})_i e^{i{\bf k(r'+r'')}} \approx \alpha '' \mu '^2 C(\omega_i)
\ee
Different models lead to different results for the function $C(\omega_i)$. Typical is a behaviour $C(\omega_i)\propto \omega_i^n$ with $n=1\ $ \cite{malinovsky} or $n=2\ $\cite{martin}. Also noninteger values for $n\ $ \cite{bermejo} and a nonvanishing limit for $\omega_i\to 0\ $ \cite{fontana} have been reported.

Using eq.(\ref{lightvib}), the sum over the different modes $i$ in (\ref{bvont}) becomes the integral
\be \label{integral}
B^{harm}(t)=\frac{2}{15} E^2 \int d\omega_i\  \alpha ''\mu '^2  g(\omega_i) C(\omega_i) f^{harm}_i(t,t_p) 
\ee
for the harmonic term. We inserted the average value $\lg P_1P_1P_2 \rg_\vartheta =2/15$. The anharmonic contribution is calculated in the same way by replacing $\alpha ''$ with $\Theta_3 \omega_i^2 \alpha '$ and $f^{harm}_i(t,t_p)$ with $f^{anharm}_i(t,t_p)$ in eq.(\ref{integral}). Instead of the time-dependent functions $f_i(t,t_p)$, the Fourier transforms $f_i(\omega,t_p)$ can be used in the integration, leading to the Fourier transformed birefringence function $B(\omega)$.

The product $g(\omega_i)C(\omega_i)\propto \omega_i^{(m+n)}$ plays the role of an effective DOS in the integral (\ref{integral}). Its frequency-dependence governs the weight of different parts of the frequency spectrum. The influence of different values is small as long as the sum $(m+n)$ is not too small $(m+n<2)$ or too large $(m+n>7)$. In these extreme cases, the frequency dependence of the effective DOS is stronger than the functional dependence on $\omega$ in the dielectric loss appearing in the functions $f_i(t,t_p)$. That means that only the low-frequency components determine the value of the integral for small $(m+n)$, and only high-frequency components determine the integral for large $(m+n)$. 

It is important to note that approximated expressions like (\ref{approx}) cannot be used in the integration (\ref{integral}), since the neglection of the damping in the denominators produces essential singularities in the integrand.

Figure 4 shows the harmonic (upper panel) and anharmonic (lower panel) contributions to the response. We used a Debye DOS, $g(\omega_i)\propto \omega_i^2, \omega_i<\omega_D=1\ $, a light to vibration-coupling $C(\omega_i)= \omega_i^2$, and a quadratically frequency-dependent damping, $\gamma(\omega_i)=0.1\omega_i^2$. For this choice of parameters, all modes are underdamped. One cycle of the pump field has been applied, $t_p=2\pi/\Omega\ $ (full lines). Dashed curves in the upper panel show results in the limit $t_p\to\infty$, that is an infinite number of pump cycles. The different curves are for three different pump frequencies $\Omega$. The most apparent difference to the individual modes' response is that only one peak is left in contrast to the two (harmonic term) or three (anharmonic term) separated signals at $\omega=\gamma_i, \omega_i, 2\omega_i$. This is because the peak at $\omega=\gamma_i$ dominates the other contributions after the integration (\ref{integral}). The positions  $\omega=2\omega_i$ (upper panel) and $\omega=\omega_i$ (lower panel) are marked by arrows. Only the anharmonic term shows small additional features at these frequencies. 

The same observation is made for other choices of the parameters. If the damping is linear in eigenfrequencies ($\gamma(\omega_i)=\kappa \omega_i$), all modes are underdamped (roughly $\kappa < 0.1$), $C(\omega_i)=\omega_i$ and the limit of large pump times is considered ($t_p\to \infty$), an analytical expression for the integral (\ref{integral}) can be obtained.  Also for this choice of parameters we observe that only one peak (at $\omega =\gamma_i$) remains in the spectrum, whereas the other contributions become negligibly small.  We do not show the result, because the expression is lengthy and not instructive, but it shows the correctness of our numerical integrations. The finding that the signal appears mainly at the damping is in contrast to our calculations for nonresonant holeburning \cite{unsers}, where the main signals appeared at the eigenfrequency. 

We would like to draw the reader's attention to the shifts of the extrema positions in figure 4. This is a consequence of frequency-selective behaviour. All modes in the chosen distribution can be treated as strongly underdamped modes. As mentioned in the individual modes' discussion, most energy is absorbed by oscillators for which the pump frequency roughly equals the eigenfrequency, $\Omega \approx \omega_i$. As the signal of these modes appears at frequency $\omega=\gamma_i(\omega_i)$ in $B^{harm}(\omega)$ and in $B^{anharm}(\omega)$, we find the maximum amplitude at $\omega \approx \gamma_i(\Omega)$, that is $\omega \approx 0.1 \Omega^2$ in the case shown in the figure. These values are marked by the dotted vertical lines. The coincidence of the maximum position with $\gamma_i(\Omega)$ holds exactly for the curves with infinite pump time $t_p$ (dashed lines in the upper panel). The extrema in case of a single applied cycle (full lines) are slightly shifted towards smaller frequencies. Apart from that small difference, the effect of increasing the number of applied cycles is mainly a growth of amplitudes.

A systematic plot of this observation is shown in figure 5. Here, the observed position of the maximum is shown as a function of the pump frequency. Plotted are results for a linear as well as a quadratic dependence of the damping as a function of $\omega_i$ for both, harmonic and anharmonic contributions. In agreement with the arguments given above, the curves exactly reproduce the assumed frequency-dependence of the damping.

We finally discuss the separation of the nonlinear $\alpha\mu\mu$-contributions from the quasi-linear signal $\propto \alpha ' \alpha'$. A phase-cycling in order to separate this contribution from the other two terms is not possible due to the fact that all terms are of order $E^2$. This has to be contrasted to the situation in NHB where a phase cycle can be devised that allows the separation of the nonlinear signal from the total response\cite{roland}. For the Kerr effect, we can either consider systems having a large dipole moment compared to the polarizability, so that the $\alpha\mu\mu$-contributions are of more importance than the $\alpha\alpha$-term. On the other hand, in a distribution of eigenmodes, the signal of the $\alpha\alpha$-term is located at the eigenfrequency $\omega \approx \omega_i$ (underdamped case, see Appendix A) in contrast to the other terms' signal at $\omega\approx\gamma_i$. Therefore, the signals should be well separated even if the individual contributions cannot be extracted. Note that this latter argument is not valid for overdamped modes. Here, extremum positions of the terms are located very close to each other.

\section*{IV. Conclusions}
We have calculated the Kerr effect response of an ensemble of independent Brownian oscillators after a sinusoidal electric field has been applied to a (glassy) sample in thermal equilibrium. Any coupling between the different modes has been neglected, and we assumed Ohmic friction for all modes. In principle, it is possible to include coupling effects as well as time-dependent damping constants, but we do not expect a qualitative change of our results as a consequence of any of these extensions. Our model is predominantly designed for the vibrational dynamics around the boson peak in disordered solids, that is the Terahertz-range. Permanent dipole moment interaction ($\mu$) is taken into account as well as induced dipole moment interaction ($\alpha$). 

From estimates of the quantities involved we conclude that the dominating terms in the signals are proportional to $\alpha ' \alpha '$, $\alpha ''\mu '\mu '$ and $\Theta_3 \alpha '\mu '\mu '$ with the cubic anharmonicity $\Theta_3$. Since we are mainly interested in the nonlinear contributions of order $\alpha\mu\mu$, we discussed the latter two terms, denoted as harmonic and anharmonic terms, in detail.

All terms were calculated in the classical BO-model, treating the external fields and the anharmonicities as perturbations. We considered the Fourier transform of the time-dependent responses, more precisely its imaginary part, since this representation turned out to be best suited for our discussion. The results contain both limiting cases, the overdamped case and the underdamped case.

For a single underdamped mode we find signals of similar amplitude at frequencies $\omega=\omega_i, 2\omega_i, \gamma_i$ with the modes eigenfrequency $\omega_i$ and its damping constant $\gamma_i$ for the harmonic and anharmonic terms, while peaks occur at $\omega=\Lambda_i, 2\Lambda_i$ with $\Lambda_i=\omega_i^2/\gamma_i$ in the overdamped case. In addition, we obtain proportionalities to dielectric loss functions evaluated at the pump frequency, $\epsilon ''(\Omega)$, or twice the pump frequency, $\epsilon ''(2\Omega)$, or at least functions closely related to the usual dielectric loss. These functions, together with the prefactor $E^2$ (the external field amplitude), substantially represent the energy absorbed by the system. It is the appearance of these functions which guarantees frequency-selective behaviour if a macroscopic system is considered.

We model a macroscopic system via a distribution of modes according to a DOS $g(\omega_i)$. We used a Debye model $g(\omega_i)\propto \omega_i^2$ with an abrupt cutoff at $\omega_i=\omega_D$. The choice of a more realistic form for the DOS like the inclusion of a bosonpeak in any form has very little influence on our results. Before being able to calculate the response of an ensemble of modes, we have to specify additional frequency-dependencies of parameters which are irrelevant for the response of an individual mode. These are the frequency-dependence of the spatial correlations appearing as prefactors in the response functions of a single mode, and the frequency-dependence of the damping, being a parameter in the individual response. The former are determined by the so-called light to vibration coupling $C(\omega_i)$ in case of the polarizability correlations, and we assume the same behaviour for the dipole moment correlations. For our model calculations, we used a dependency $C(\omega_i)=\omega_i^2$. As above, results for e.g. $C(\omega_i)=\omega_i$ hardly differ from the case discussed. For the damping, we assumed a proportionality $\gamma_i(\omega_i)\propto \omega_i^{\eta}$. Our main finding is that any functional form $\gamma_i(\omega_i)$, e.g. $\eta =2$, can be extracted from experimental Kerr effect data. Due to the frequency-selectivity mainly those modes with $\omega_i \approx \Omega$ (in the strongly underdamped case) contribute to the signal. We further find that the signal of a distribution of modes mainly shows a peak at the damping ($\omega=\gamma_i$) of these modes, whereas the contributions at $\omega=\omega_i$ and at $\omega=2\omega_i$, clearly visible in the individual modes' response, vanish. As a consequence, the signal shows a maximum at $\omega \approx \gamma_i(\Omega)$. Therefore, if the time-dependent evolution of the system's polarizability after an excitation with a sinusoidal field of pump frequency $\Omega$ is measured for several pump frequencies, the signals are Fourier transformed, and the maxima positions are plotted versus $\Omega$, the frequency-dependence of the damping, $\gamma_i(\omega_i)$, can be determined from the data. We think it would be interesting to compare such results to data from other experimental techniques capable to detect the damping's frequency dependence, like neutron scattering or inelastic X-ray scattering.

Although our discussion is focussed on the capability to extract the damping's frequency-dependence, our results show that additional information about the dynamics is contained in the Kerr effect signals, for example information about the origin of the nonlinear signal like anharmonicity or higher-order derivatives of the polarizability. The principle finding of frequency-selective behaviour allows to specifically address certain dynamic features in the sample, which is the main advantage of nonlinear techniques as compared to linear spectroscopy.

\section*{Acknowledgement}
This work has been supported by the DFG under contract No. Di693/1-2. 

\newpage
\begin{appendix}
\section*{Appendix A}
\setcounter{equation}{0}
\renewcommand{\theequation}{A.\arabic{equation}}

\subsubsection*{Full expressions for all contributions}
The exact solutions for the harmonic and anharmonic terms (\ref{bharm},\ref{banharm}) are
\Be \nonumber
f^{harm}_i(t,t_p) & = &\frac{1}{(\l_1-\l_2)^2} \sum_{(i,j)=1,2} (-)^{i+j} e^{-(\l_i+\l_j)t}\  \frac{\Omega^2}{\l_i^2+\Omega^2} \Bigg[ \frac{1}{\l_j^2+\Omega^2} \left( e^{-(\l_i+\l_j)t_p}-e^{-\l_i t_p} \right) \\ 
& & \quad - \frac{1}{(\l_i+\l_j)^2+4\Omega^2}\left( 1+\frac{2\l_i}{\l_i+\l_j} \right) \left( e^{-(\l_i+\l_j)t_p}-1  \right) \Bigg] \\ \nonumber
f^{anharm}_i(t,t_p) & = & \frac{-2}{(\l_1-\l_2)^3} \sum_{(i,j,k)=1,2} (-)^{i+j+k} \frac{1}{\l_i-\l_j-\l_k} \frac{\Omega^2}{\l_k^2+\Omega^2}     \\ \nonumber
& & \times \Bigg[ e^{-(\l_k+\l_j)t} \frac{1}{(\l_k+\l_j)(\l_j^2+\Omega^2)\left((\l_k+\l_j)^2+4\Omega^2 \right)} \Big\{ (3\l_k+\l_j)(\l_j^2+\Omega^2)  \\ \nonumber
& & \quad - e^{-\l_k t_p}(\l_k+\l_j)((\l_k+\l_j)^2+4\Omega^2)-e^{-(\l_k+\l_j)t_p}(\l_k^2+\Omega^2)(-\l_k-3\l_j)\Big\}  \\ \nonumber
& & - e^{-\l_i t} \frac{1}{\l_i\left((\l_i-\l_k)^2+\Omega^2 \right) (\l_i^2+4\Omega^2)} \Big\{ (2\l_k+\l_i)\left( (\l_i-\l_k)^2+\Omega^2 \right) \\ 
& & \quad - e^{-\l_k t_p}\l_i(\l_i^2+4\Omega^2)-e^{-\l_i t_p}(\l_k^2+\Omega^2)(2\l_k-3\l_i) \Big\} \Bigg] 
\Ee
with the eigenvalues $\lambda_i$ defined after eq.(\ref{fharmlimit}). In the summations, all indices take on values $1$ and $2$.

The term $\propto \alpha '\alpha '$ is
\begin{equation}
B^{ql}_i(t)=E^2 \langle P_2P_2 \rangle_\vartheta \int d^3{\bf r}  \int d^3{\bf r'} \alpha'({\bf r})_i\alpha'({\bf r'})_i  e^{i{\bf k r'}}\ f^{ql}_i(t,t_p)
\end{equation}
with
\be
f^{ql}_i(t,t_p)=\frac{2}{\l_1-\l_2}\left(\frac{1}{\l_2} \frac{\Omega^2}{\l_2^2+4\Omega^2}(1-e^{-\l_2 t_p})e^{-\l_2 t} - \frac{1}{\l_1} \frac{\Omega^2}{\l_1^2+4\Omega^2}(1-e^{-\l_1 t_p})e^{-\l_1 t} \right) 
\ee
which in the overdamped limit becomes
\be \label{quasilinoverdamp}
f^{ql, OU}_i(t,t_p)=\frac{\Omega}{\omega_0^2 \Lambda_i} \epsilon ''(2\Omega) (1-e^{-\Lambda_i t_p}) e^{-\Lambda_i t}
\ee
Here, the superscript 'ql' stands for 'quasi-linear'. The numerical value of the orientational average is $\lg P_2P_2 \rg_\vartheta=1/5$. The maximum amplitude of this term is achieved for $\Omega \approx \omega_i/2$ in the strongly underdamped case and $\Omega \approx \Lambda_i/2$ in the overdamped case.

Fourier transforms $f_i(\omega ,t_p)$ for all the functions $f_i(t,t_p)$ can be derived from the given expressions by $FT(e^{-\xi t})=\int_0^\infty dt e^{i\omega t}e^{-\xi t}=1/(\xi -i \omega)$, for $\xi =\Lambda_i, \lambda_1, \lambda_2$ or a linear combination of the eigenvalues.

\section*{Appendix B}

\setcounter{equation}{0}
\renewcommand{\theequation}{B.\arabic{equation}}

\subsubsection*{Reorientational models}
If reorientational models are considered we can skip the spatial dependence of $\alpha$ in the expectation value (\ref{definebs}). Here, the number $\alpha$ appears as a prefactor and the influence of external fields is a modification of the orientational distribution function (isotropic in the equilibrium state). We therefore write the time-dependent polarizability as
\be \label{rotdefineb}
\alpha \  \lg P_2(\cos \vartheta(t)) \ \rg =B^{rot}(t)
\ee
in contrast to eq.(\ref{definebs}). Note that now the orientations are the time-dependent part.

The expectation value $\langle P_2(\vartheta (t)) \rangle$ is calculated using some dynamic model for the orientations, e.g. with the rotational diffusion equation \cite{kubo}
\begin{equation}\label{rotdiff}
\dot P(\Omega ,t| \Omega ')=D \left[ \frac{1}{\sin \vartheta}\partial_\vartheta \left[ \sin \vartheta \left(\partial_\vartheta  + \beta \frac{\partial {\cal H}_{ext}(\vartheta ,t)}{\partial \vartheta}\right) \right] +\frac{1}{\sin^2 \vartheta} \partial^2_\phi \right] P(\Omega ,t|\Omega ') 
\end{equation}
This equation is the Fokker-Planck-equation for the conditional probability $P(\Omega ,t|\Omega ')$ to find the orientation $\Omega$ at time $t$, assumed the orientation was $\Omega '$ at time $0$. $\vartheta$ and $\phi$ denote Eulerian angles, $\beta =(k_B T)^{-1}$ and $D$ is the rotational diffusion constant. The external Hamiltonian is again given by eq.(\ref{exthamilton}). For the solution of eq.(\ref{rotdiff}), the propagator $P(\Omega,t|\Omega ')$ is expanded in a series of Wigner matrices. All details about this expansion and helpful orthogonality relations of the Wigner matrices can be found in \cite{hansgregor}. 

As in the BO-model, the interaction with the external field is calculated using perturbation theory. Note that the corresponding operator in the Fokker-Planck-equation is proportional to $\frac{\partial {\cal H}_{ext}}{\partial \vartheta}$ instead of $\frac{\partial {\cal H}_{ext}}{\partial q_i}$ as in the BO-model. Therefore no derivatives of permanent and induced dipole moment ($\mu^{(n)}, \alpha^{(n)}$) appear in the rotational diffusion model. Instead, the perturbation terms are proportional to $\alpha$ and $\mu$, and to derivatives of the corresponding Legendre polynomial $P_1$ or $P_2$ with respect to $\vartheta$. 

The linear response of order $E$ in $B^{rot}(t)$ is calculated with the permanent dipole interaction of order $\mu E$ as a first order perturbation. As for the BO-model, this linear response vanishes. The nonvanishing contributions of order $E^2$ are proportional to $\alpha \alpha$ and to $\alpha \mu \mu$, where the former corresponds to the quasi-linear response $\propto \alpha '\alpha '$ in the BO-model, and the latter corresponds to both harmonic and anharmonic terms (see above) in the BO-model. We find the Kerr effect reponse
\Be \label{langerausdruck} \nonumber
B^{rot}(t) & =  & \frac{1}{30} E^2\alpha \beta e^{-\Gamma_2 t} \Bigg\{  \alpha  \frac{\Omega}{D} \epsilon ''(\Gamma_2,2\Omega) \left[1-e^{-\Gamma_2 t_p}\right] \\ \nonumber
& & + \beta \mu^2  \epsilon ''(\Gamma_1 ,\Omega) \bigg( 3 \epsilon ''(\Gamma_2-\Gamma_1 ,\Omega) \left[e^{-\Gamma_2 t_p}-e^{-\Gamma_1 t_p}\right] \\
& & \qquad \qquad \qquad + \frac{5}{3} \epsilon ''(\Gamma_2, 2\Omega) \left[1-e^{-\Gamma_2 t_p}\right]   \bigg)    \bigg\}
\Ee
The dielectric loss is defined as $\epsilon ''(x,y)=\frac{xy}{x^2+y^2}$ and $\Gamma_l=l(l+1) D$ in the rotational diffusion model. Note that the time-dependence is a monoexponential decay with the rate $\Gamma_2=6 D$, i.e. the decay rate of the second Legendre polynomial, cf. ref.\cite{williams}. If we consider the limit of many pump cycles or large $t_p$ (that means $N\gg \Omega/(4\pi D)$ in this model) we find
\be \label{rotdiffresp}
B^{rot}(t,t_p=\infty )=\frac{1}{6} E^2\alpha \beta e^{-6 D t} \epsilon ''(3D,\Omega) \left(\frac{1}{5} \alpha \frac{\Omega}{D} + \frac{1}{3} \beta \mu^2 \epsilon ''(2D,\Omega) \right) 
\ee
The situation is very similar to the overdamped responses in the BO-model, see eqs.(\ref{ouharm},\ref{ouanharm},\ref{quasilinoverdamp}). For a description of a macroscopic system, one would choose a distribution of rotational diffusion constants. Usually, such a distribution is chosen in a manner that the linear dielectric response has the stretched exponential form which is often observed in experiments on glassy materials.
As a consequence of the appearance of the dielectric loss functions in eq.(\ref{langerausdruck})), most of the energy in a distribution is absorbed for $\Omega \approx \Gamma_1$ and $2\Omega\approx\Gamma_2$, respectively for $\Omega \approx (2\dots 3) D$. We therefore have for the product $\Gamma_l t_p = l(l+1)D2\pi N/\Omega \approx 2\pi N$ for the mainly energy absorbing modes. Therefore, the appearing exponentials $\exp(-\Gamma_l t_p)$ in eq.(\ref{langerausdruck}) are practically zero after only one applied cycle, meaning that the limit (\ref{rotdiffresp}) is always a good approximation, very similar as for the overdamped BO-model response.  

Instead of rotational diffusion, other reorientational models can also be considered, for example isotropic or anisotropic random jump models. We expect a similar form of the Kerr effect response also for these models. 

\end{appendix}

\newpage
\subsection*{Figure captions}
\begin{description}
\item[Fig.1 : ]  Electric field sequence: One or several full cycles of the form $E(\tau)=E\sin{(\Om \tau)}$ excite a sample in thermal equilibrium. In the text, $\Om$ is denoted as 'pump frequency' and $t_p\!=\!2N\pi/\Om$ as 'pump time' with $N$ denoting the number of cycles. After the field is switched off at $t_p$, the time-dependent evolution of the polarizability is monitored.
\item[Fig.2 : ] Imaginary part of the Fourier transformed response of a single underdamped mode for the harmonic term (upper panel) and anharmonic term (lower panel). The eigenfrequency is $\omega_i=1$ and the damping $\gamma_i=0.1$. The pump frequency is $\Omega=1$. For the harmonic term responses of one (full line) and an infinite number (dashed line) of applied cycles are shown. Because the influence of increasing the pump time is mainly a growth of amplitudes we show only the response of one applied cycle for the anharmonic term (lower panel). The features' positions $\omega=\gamma_i,\omega_i,2\omega_i$ are marked by vertical lines.
\item[Fig.3 : ] As in figure 2 for an overdamped mode. Here is $\omega_i=1$ and $\gamma_i=10$, leading to a Ornstein-Uhlenbeck rate of $\Lambda_i=0.1$. The pump frequency is $\Omega=0.1$. One cycle has been applied.
\item[Fig.4 : ] Upper panel: Imaginary part of the Fourier transformed signal $[B^{harm}(\omega)]''$ from a distribution of modes (harmonic term). The mode distribution function is a (normalized) Debye, $g(\omega_i) \propto \omega_i^2, \omega_i<1$. The light to vibration coupling is $C(\omega_i)=\omega_i^2$, and the damping is chosen as $\gamma_i(\omega_i)=0.1\ \omega_i^2$. Shown are results for three different pump frequencies $\Omega=0.01, 0.03, 0.1$ from left to right, in each case for one applied cycle, $N=1$, (full lines), and for an infinite number of cycles, $N\to\infty$ (dashed lines). The frequencies $\omega=\gamma_i(\Omega)$ are marked by vertical dotted lines (extrema positions of the corresponding responses), and marked by arrows at $\omega=2\Omega$. No features are visible at the latter positions. The prefactors $2/15\ E^2 \alpha ''\mu '^2$ were set to unity. \\ Lower panel: As in the upper panel for the anharmonic term, $[B^{anharm}(\omega)]''$. Here, minor features are observable at $\omega=\Omega$ (marked by arrows) and at $\omega=2\Omega$. Only results for one cycle are shown. Note that amplitudes are plotted negative. Prefactors set to unity are $2/15\ E^2\Theta_3 \alpha '\mu '^2$.
\item[Fig.5 : ] The extrema positions of $[B^{harm}(\omega)]''$ and $[B^{anharm}(\omega)]''$, plotted versus pump frequency $\Omega$. DOS and light to vibration coupling are chosen as in figure 4. The upper line is for frequency-dependent damping $\gamma_i(\omega_i)=0.1\ \omega_i$, the lower lines for $\gamma_i(\omega_i)=0.1\ \omega_i^2$. The functions exactly reproduce the assumed frequency-dependence of the damping. Results of harmonic and anharmonic terms are not distinguishable. Full lines are for one cycle of the pump field. The dashed line stems from the harmonic term in the limit of large pump times, illustrating the slight shift towards larger $\omega$ with increasing $t_p$ mentioned in the discussion of figure 4.  
\end{description}

\newpage
\pagestyle{empty}
\begin{figure}
\includegraphics[width=15cm]{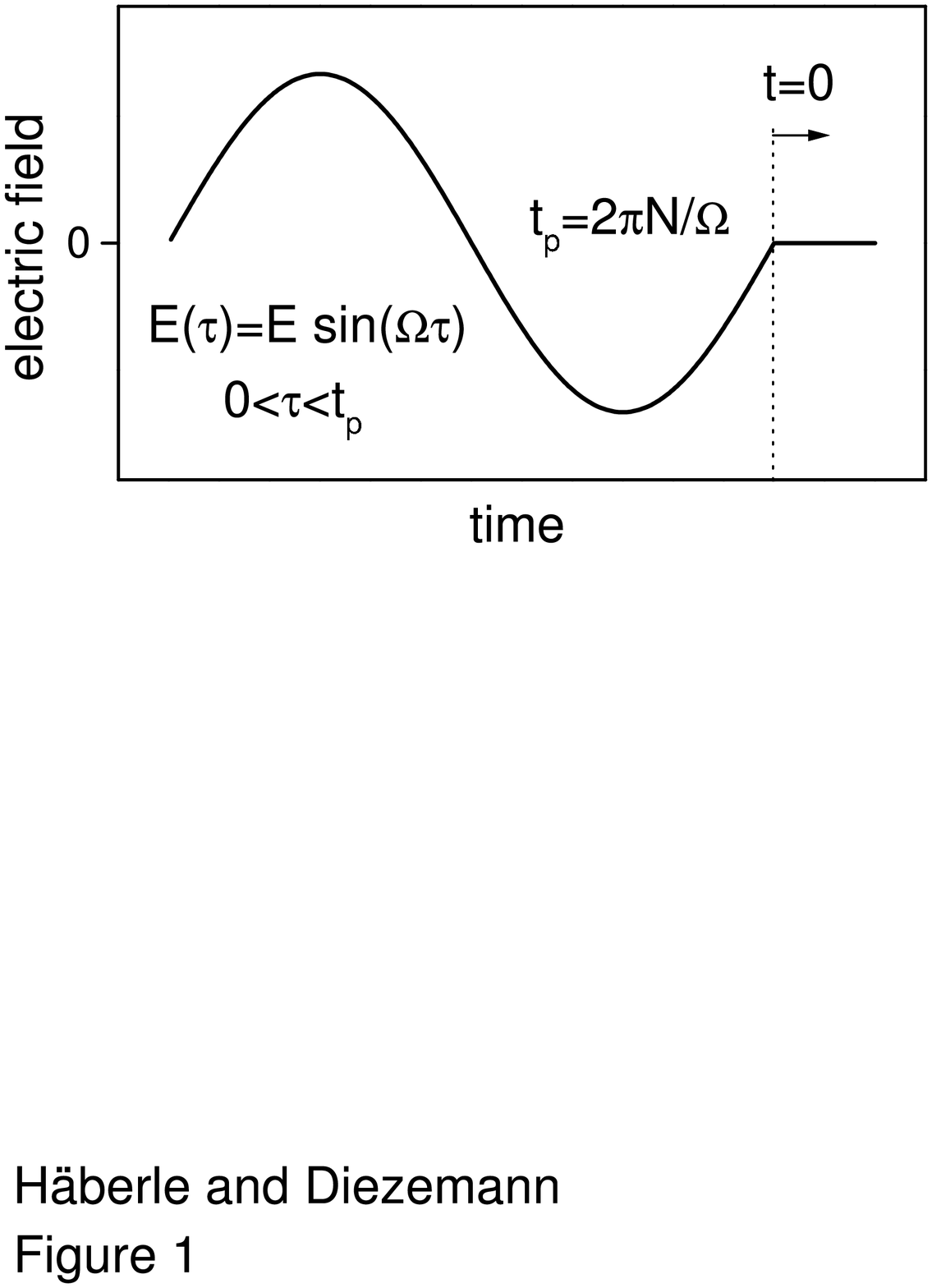}
\end{figure}

\newpage
\pagestyle{empty}
\begin{figure}
\includegraphics[width=15cm]{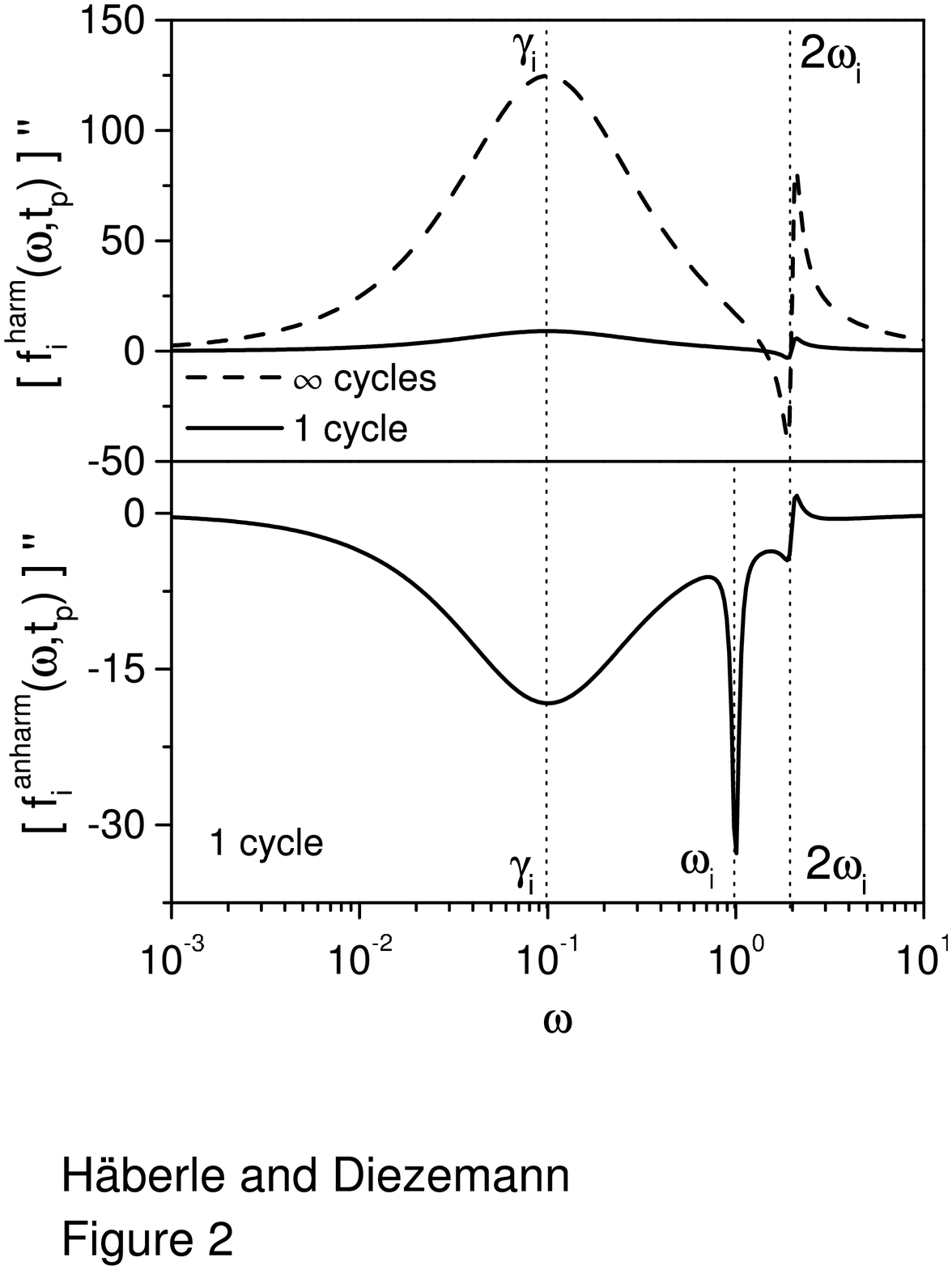}
\end{figure}

\newpage
\pagestyle{empty}
\begin{figure}
\includegraphics[width=15cm]{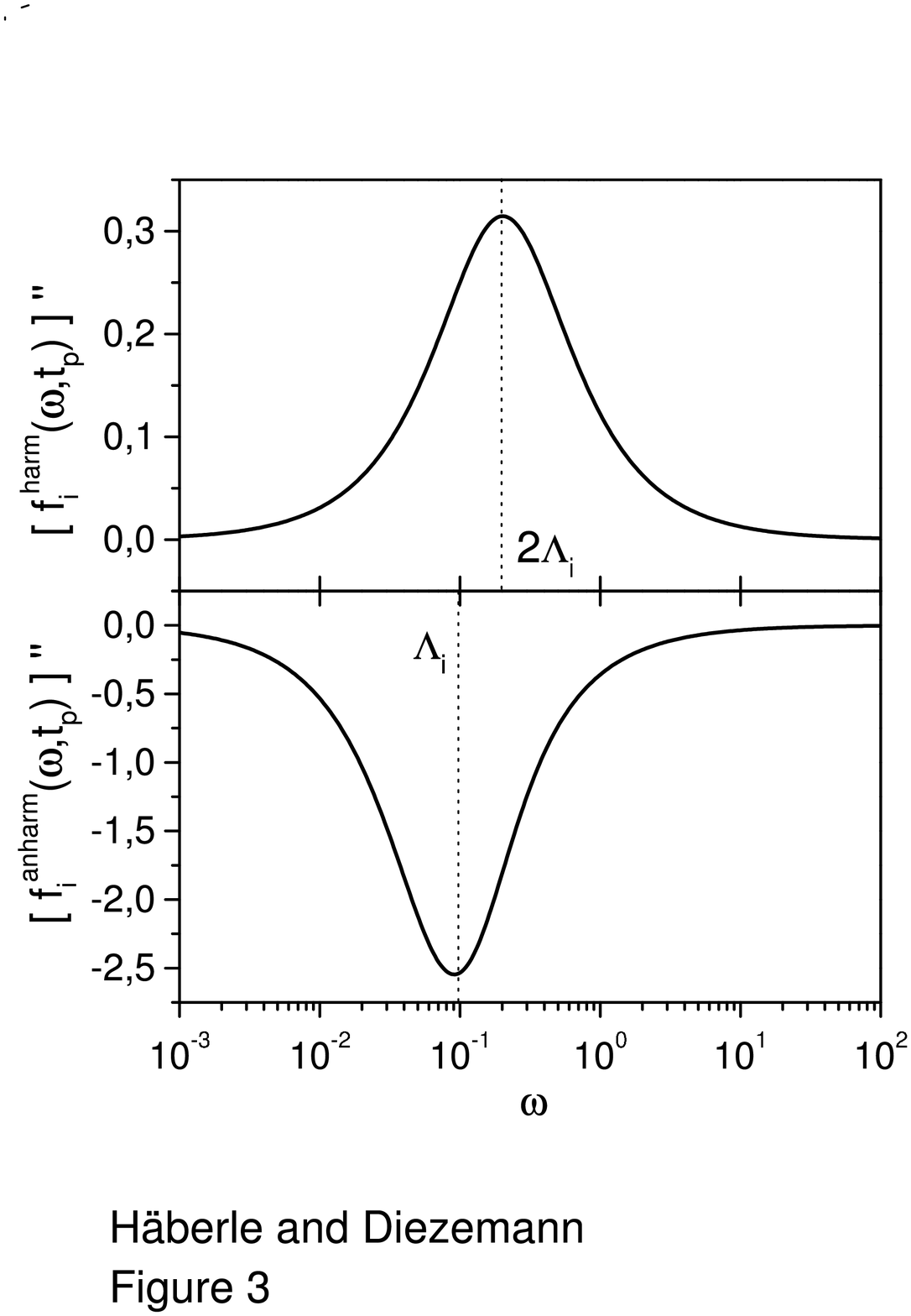}
\end{figure}

\newpage
\pagestyle{empty}
\begin{figure}
\includegraphics[width=15cm]{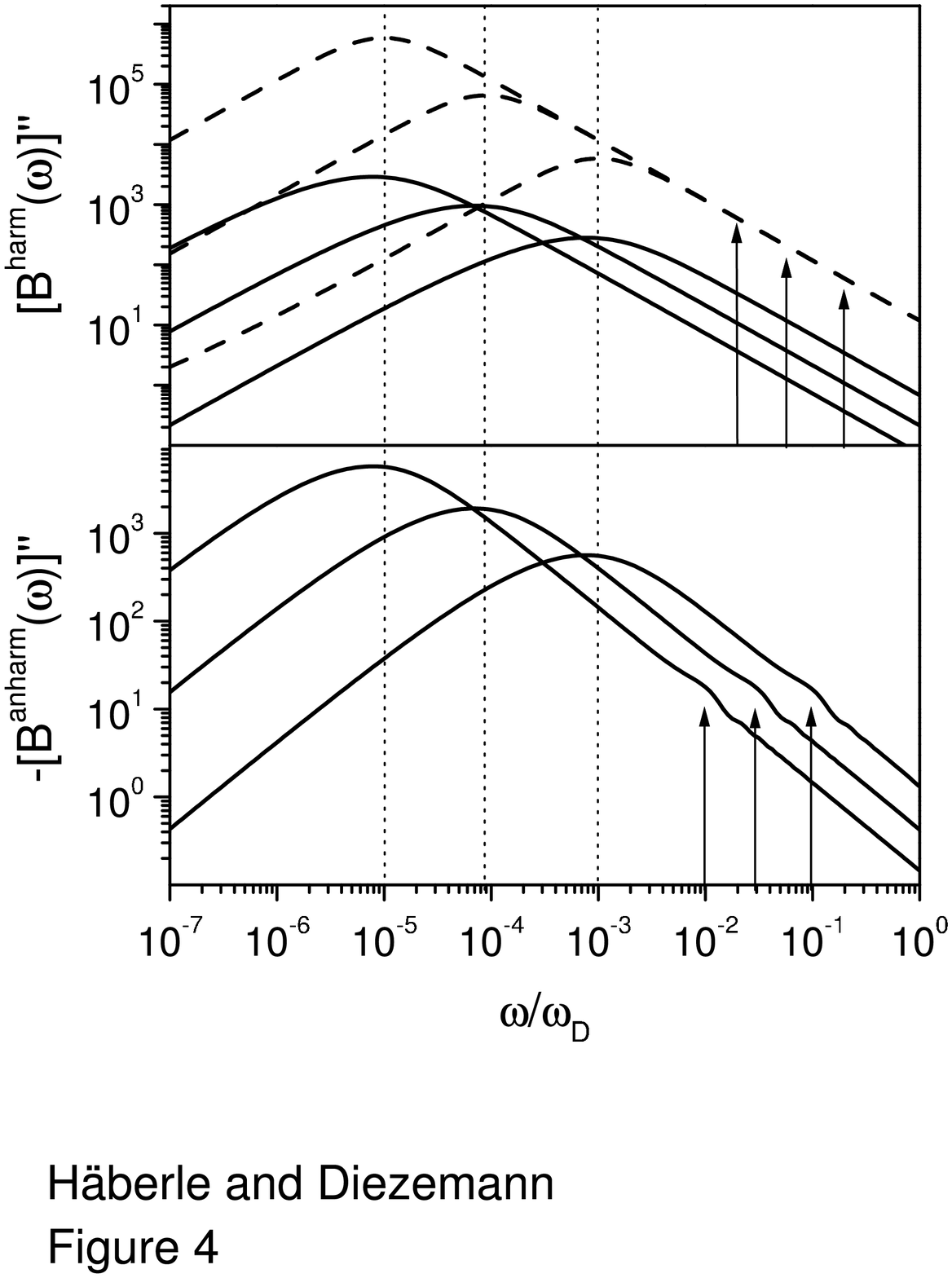}
\end{figure}

\newpage
\pagestyle{empty}
\begin{figure}
\includegraphics[width=15cm]{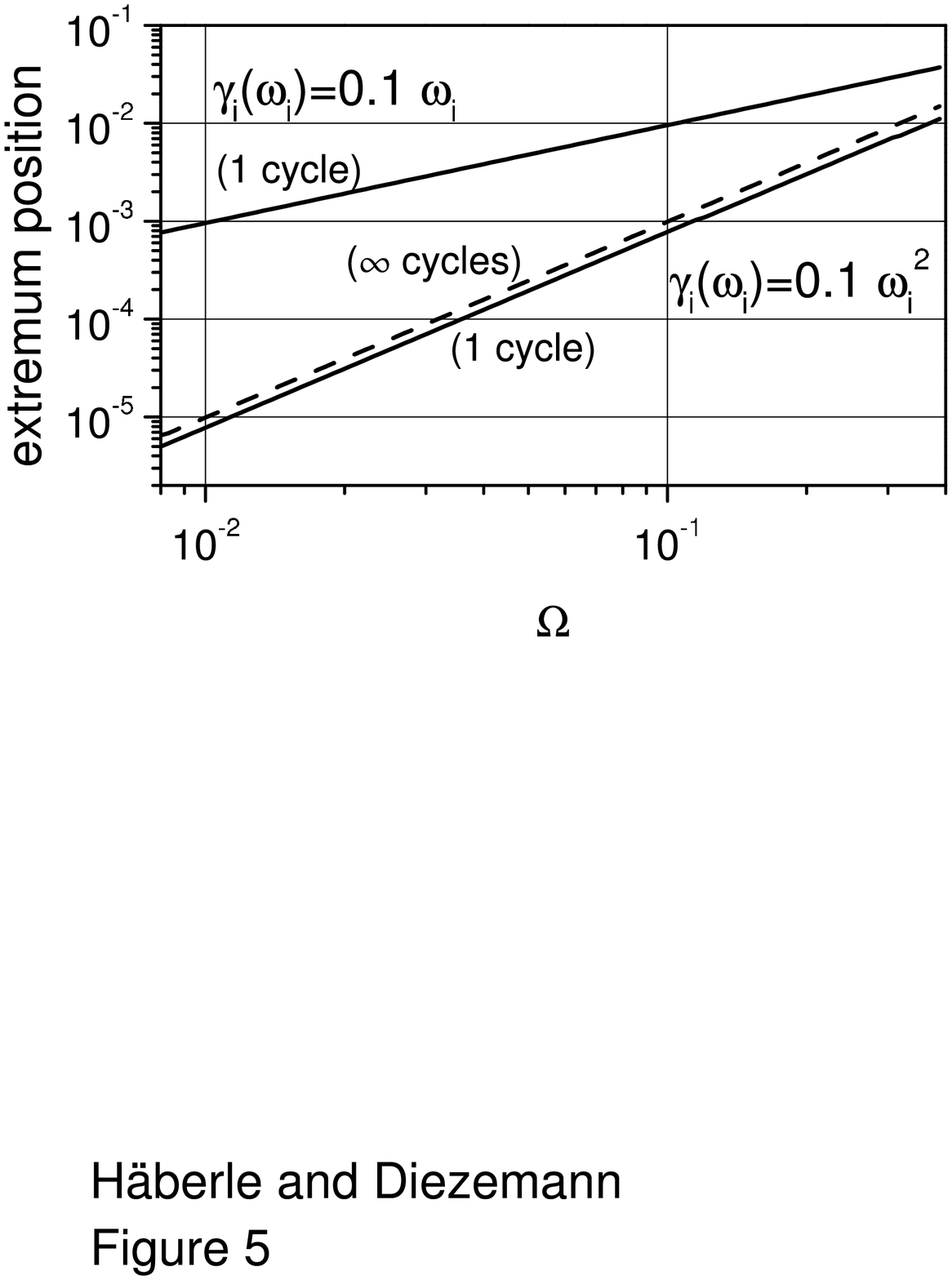}
\end{figure}


\begin{thebibliography}{XX}
\bibitem{meinsenf}
Note that we collect terms of equivalent order in the exciting field. Alternatively, the exponents of the field can be ignored and equivalent orders of the perturbation can be considered. From that point of view, the terms $\propto \alpha\alpha$ and (vanishing in isotropic systems) $\propto \alpha\mu$ could be considered as first order, while the terms $\propto \alpha\mu\mu$ and $\propto \alpha\alpha\alpha$ were of second order. The difference arises from the fact that the forces in the equation of motion are $\propto E$ in the permanent dipole interaction, and $\propto E^2$ in the induced dipole interaction.
\end{thebibliography}
\end{document}